\documentclass[aps,amsfonts,amsmath,amssymb,prl,showpacs]{revtex4}

\usepackage{graphicx}
\newcommand{\be}{\begin{equation}}
\newcommand{\ee}{\end{equation}}
\newcommand{\bex}{\begin{eqnarray}}
\newcommand{\eex}{\end{eqnarray}}
\newcommand{\bsub}{\begin{subequations}}
\newcommand{\ensub}{\end{subequations}}
\newtheorem{thm}{Theorem}

\def\qed{$\Box$}

\newcommand{\ou}{\rm ~OR~}
\newcommand{\og}{\rm ~AND~}
\newcommand{\ouh}{\rm ~OR}

\begin{document}
\title{Assisted Quantum Secret Sharing}
\author{Sudhir Kumar Singh}
\thanks{Moving to Dept. of Electrical Engg., Univ. of California, Los Angeles.}
\affiliation{Dept. of Mathematics, Indian Institute of Technology,
Kharagpur- 721302, India}
\author{R. Srikanth}
\email{srik@rri.res.in}
\affiliation{Optics Group, Raman Research Institute, Bangalore- 560080, India}
\begin{abstract}
A restriction on quantum secret sharing (QSS) that comes from 
the no-cloning theorem is that any pair of authorized sets in an access structure should overlap. 
From the viewpoint of application, this places an unnatural constraint on secret sharing.
We present a generalization, called assisted QSS (AQSS), where access structures 
without pairwise overlap of
authorized sets is permissible, provided some shares are withheld by the share dealer.
We show that no more than $\lambda-1$ withheld shares are required, where $\lambda$ is
the minimum number of {\em partially linked classes} among the authorized sets for the QSS.
This is useful in QSS schemes where the share dealer is honest by definition and is
equivalent to a secret reconstructor. Our result means that such applications of QSS 
need not be thwarted by the no-cloning theorem.
\end{abstract}

\maketitle


Suppose the president of a bank, Alice, wants to give access to a vault to two 
vice-presidents, Bob and Charlie, whom she does not entirely trust.
Instead of giving the combination to any one of them, she may desire to
distribute the information in such a way that no vice-president alone has any 
knowledge of the combination, but both of them can jointly determine the combination.
Cryptography provides the answer to this question in the form of
{\it secret sharing} \cite{schneier96}.
In this scheme, some sensitive data is distributed among a
number of parties such that certain authorized sets of parties can 
access the data, but no other combination of players.
A particularly symmetric variety of secret splitting (sharing) is
called a {\it threshold scheme}: 
in a $(k,n)$ classical threshold scheme (CTS),
the secret is split up into $n$ pieces (shares),
of which any $k$ shares form a set authorized to reconstruct the secret, while any set of 
$k-1$ or fewer shares has no information about the secret.
Blakely \cite{blakely79} and Shamir \cite{sha79} showed that CTS's exist
for all values of $k$ and $n$ with $n \geq k$. By concatenating threshold schemes,
one can construct arbitrary access structures, subject only to the condition of
monotonicity (ie., sets containing authorized sets should also be authorized)
\cite{ben90}. Hillery {\em et al.}
\cite{hil00} and Karlsson {\em et al.} \cite{kar00} proposed methods for
implementing CTSs that use {\em quantum} information
to transmit shares securely in the presence of eavesdroppers.

Subsequently, extending the above idea to the quantum case,
Cleve, Gottesman and Lo \cite{cle00}, using 
the notion of quantum erasure correction \cite{cs,gra97}, presented
a $(k,n)$ {\it quantum} threshold scheme (QTS) as a method to split up 
an unknown secret quantum state $|S\rangle$ into $n$ 
pieces (shares) with the restriction that 
$k > n/2$-- this inequality being needed to ensure that no two disjoint sets of players
should be able to reconstruct the secret, in conformance with 
the quantum no-cloning theorem \cite{woo82}.
QSS has been extended beyond QTS to
general access structures \cite{got00,smi00}, but here 
none of the authorized sets shall be mutually disjoint:
given a QSS access structure $\Gamma = \{\alpha_1,\cdots,\alpha_r\}$ over $N$
players, the no-cloning restriction entails that:
\begin{equation}
\label{noklo}
\alpha_j \cap \alpha_k \ne \phi~~~~~\forall j,k.
\end{equation}
Potential applications of QSS include
creating joint checking accounts containing quantum money \cite{wiesner83}, 
or sharing hard-to-create ancilla
states \cite{got00}, or performing secure distributed quantum computation \cite{cre01}.
A tri-qubit QSS scheme has recently been implemented \cite{lan04}.
The chances of practical implementation of QSS are improved by employing equivalent
schemes that maximize the proportion of classical information processing \cite{nas01,sud03}.

The requirement Eq. (\ref{noklo}) places a restriction quite unnatural to applications, 
where we may more likely expect to find groups of people with
mutual trust within the group, and hardly any outside it.
Our present work is aimed at studying a way to overcome this limitation.
In particular, we show that allowing the dealer to withhold a small number of shares 
permits arbitrary access structures to be acceptable, subject only to monotonicity.
This modified scheme we call ``assisted QSS" (AQSS), the shares withheld by the dealer
being called ``resident shares". While more general than conventional QSS, AQSS
is clearly not as general as classical secret sharing, since it requires 
shares given to the (non-dealer) players, called ``player shares",
to be combined with the resident shares for reconstructing the secret.

Inspite of this limitation, the modified scheme
can be useful in some applications of secret sharing,
in particular, those in which the secret dealer 
is by definition a trusted party and 
where re-construction of the secret effectively occurs by re-convergence of
shares at the dealer's station. 
In the bank example above, access is allowed by the
bank vault (which can be thought of effectively as the dealer, acting as the bank
president's proxy) if the secret reconstructed from
the vice-presidents' shares is the required password. The locker thus effectively 
serves as both the dealer and site of secret reconstruction.
In AQSS, the player shares are combined with the
resident share(s) to reconstruct the secret. Clearly, this leads to no 
loss of generality in this type of QSS. 
Where the secret dealer is not necessarily trusted, such as in
multi-party secure computation (MPSC), AQSS may be less useful, though
here again only a more detailed study can tell whether MPSC cannot be turned into a suitable
variant of AQSS.

It is assumed that all the $n$ (quantum) shares are somehow divided among the $N$
players. In an AQSS scheme, $m < n$ shares are allowed to remain with 
the share dealer, as resident shares. 
In order that AQSS should depart minimally
from conventional QSS, we further require that the number of resident shares should be {\em the
minimum possible} such that a violation of Eq. (\ref{noklo}) can be accomodated.
Thus, a conventional QSS access structure like
$\Gamma = \{ABC, ADE, BDF\}$, which as such conforms to the no-cloning theorem,
will require no share assistance. 
A conventional QSS scheme is a special case of AQSS,
in which the set of resident shares is empty. 
We prove by direct construction in the following Theorem that, by allowing for non-zero resident
shares, the restriction (\ref{noklo}) does not apply to AQSS.
Therefore, with share assistance,
the only restriction on the access structure $\Gamma$ in AQSS is monotonicity, as
with classical secret sharing.

Given access structure $\Gamma = \{\alpha_1,\cdots,\alpha_r\}$,
we divide all authorized sets $\alpha_j$ 
into {\em partially linked classes}, each of which 
is characterized by the following two properties: (a) Eq. (\ref{noklo}) is satisfied 
if $j, k$ belong to the same class; (b) for any two distinct classes, there is at least
one pair $j,k$, where $j$ belongs to one class and $k$ to the other, such that
Eq. (\ref{noklo}) fails. 

A division of $\Gamma$ into such classes we call as a {\em partial link classification}.
The number of classes in a partial link classification gives its size.
In general, neither the combinations nor size of partial link classifications
are unique. We denote the size of the {\em smallest} partial link classification for a given
$\Gamma$ by $\lambda$. If all authorized sets have mutual pairwise overlap then 
$\lambda=1$ and the single partially linked class is, uniquely, $\Gamma$ itself,
and AQSS reduces to conventional QSS. If none of the $\alpha_j$'s have mutual pairwise
overlap, then $\lambda=r$ and the $r$ partially linked classes are, uniquely, each $\alpha_j$.
If there are $s$ disconnected groups of $\alpha_j$'s (that is, Eq. (\ref{noklo})
fails for all pairs $j, k$, where $j$ comes from one group and $k$ from another)
then $\lambda \ge s$. The inequality arises from the fact that there may be more than
one partially linked class within a disconnected group.

\begin{figure}
\includegraphics{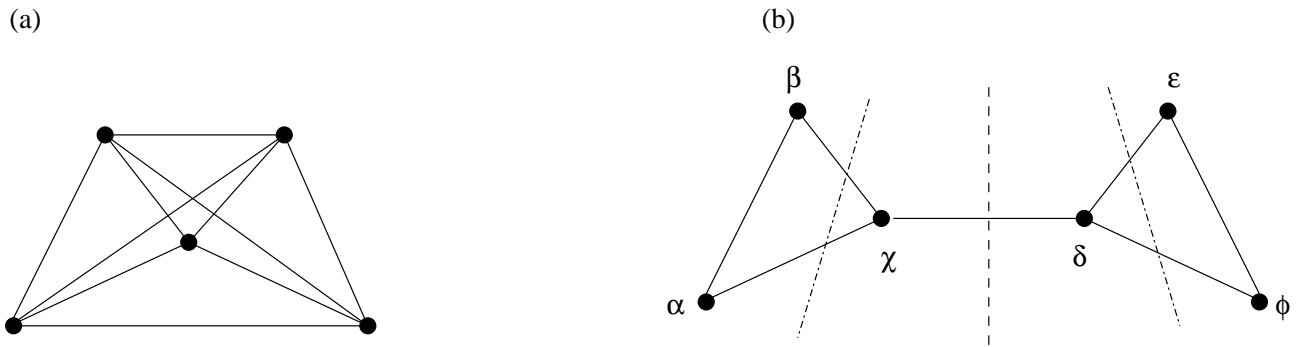}
\caption{The vertices represent authorized sets, the edges depict a non-vanishing
pairwise overlap between two authorized sets. Figure (a) represents a
conventional QSS, where all sets have pairwise overlap, meaning that the AS graph is
complete, so that $\lambda=1$. Figure (b) represents
a situation where this does not hold and hence $\lambda > 1$. The authorized sets
are labelled $\{\alpha, \beta, \chi, \epsilon, \delta, \phi\}$. The dashed line is a 
cut leading to two partially linked classes (the pair of 3-cliques, $\{\alpha,\beta,\chi\}$ and 
$\{\delta,\epsilon,\phi\}$), so that 
$\lambda=2$. The two dash-dotted lines are system of two cuts leading
to three partially linked classes (the triple of 2-cliques,
$\{\alpha,\beta\}$, $\{\chi,\delta\}$ and $\{\epsilon,\phi\}$).}
\label{AQSS}
\end{figure}

The problem of obtaining a partial link classification can be analyzed graph theoretically.
It is easy to visualize $\Gamma$ as a graph $G(V,E)$, composed of a set $V$ of vertices
and set $E$ of edges. The vertices are the authorized sets, $V=\{\alpha_j\}=\Gamma$ and 
edges $E=\{(\alpha_j,\alpha_k)\}$ correspond to pairs of sets that have pairwise overlap.
Such a graph may be called an {\em access structure graph} (AS graph) for $\Gamma$.
A partial link classification corresponds to a partitioning of the AS graph $G$ such that
each partition is a 
{\em clique}, i.e., a {\em complete} subgraph in $G$ (A graph is called complete
if its each vertex has an edge with its every other vertex). 
Figure \ref{AQSS}(a) depicts a conventional QSS, where $\Gamma$ is partitioned into
a single 5-clique. Figure \ref{AQSS}(b) depicts a
more general case covered by AQSS, where $\Gamma$ is partitioned into a pair of
3-cliques or into a triple of 2-cliques.
The problem of determining $\lambda$ is thus equivalent to the combinatorial problem
of partitioning $G$ into the minimum number of cliques. Here it is worth noting that
many multi-party problems are amenable to combinatorial treatment.

Before introducing the main Theorem, it is instructive to look at the classical situation. 
In our notation, single (double) parantheses indicate CTS (QTS).
For a classical secret sharing scheme, suppose $\Gamma = \{ABC, AD, DEF\}$,
which can be written in the normal form $\{(A \og B \og C) \ou (A \og D)
\ou (D \og E \og F)\}$. The
\og gate corresponds to a $(|\alpha_j|,|\alpha_j|)$ 
threshold scheme, while \ou to a (1,2) threshold
scheme. By concatenating these two layers, we get a construction for $\Gamma$.
In the conventional QSS, the above fails for two reasons, both connected to the
no-cloning theorem: the members of $\Gamma$ should not be disjoint; and further there
is no $((1,2))$ scheme.
However, we can replace $((1,2))$ by a $((2,3))$ scheme, which corresponds to a majority
function of \ouh. In general, we replace a $((1,r))$ scheme by a $((r,2r-1))$
scheme. $r$ of the shares correspond to individual authorized sets in $\Gamma$,
shared within an $\alpha_j$ according to a $((|\alpha_j|,|\alpha_j|))$ threshold scheme,
and, recursively, the other $r-1$ shares are shared according to a pure state
scheme that implements a maximal structure $\Gamma_{\max}$ that includes
$\Gamma$ (obtained by adding authorized sets to $\Gamma$ until the complement of
every unauthorized set is precisely an authorized set) \cite{got00}. The Theorem below 
extends this idea to the situation where $\Gamma$
does not satisfy Eq. (\ref{noklo}). 
\begin{thm}
Given an access structure $\Gamma = \{\alpha_1, \alpha_2,
\cdots,\alpha_r\}$ with a minimum of $\lambda$ partially linked classes
among a set of players ${\cal P} = \{P_1, P_2, \cdots, P_N\}$,
an assisted quantum secret sharing scheme exists iff $\Gamma$ is monotone. It requires
no more than $\lambda-1$ resident shares.
\end{thm}
\noindent
{\bf Proof:} 
We give a proof by construction. It is known that if $\lambda=1$, then there exists a 
conventional QSS to realize it \cite{got00}.
Suppose $\lambda>1$. To implement $\Gamma$ (which represents a monotonic
access structure), the dealer first employs a
$((\lambda,2\lambda-1))$ majority function, assigning one share to each class.
Recursively, each share is then subjected to a conventional QSS within each
class. The remaining $\lambda-1$ shares remain resident with the dealer. To reconstruct
the secret, any authorized set can reconstruct the share assigned to its class,
which, combined with the resident shares, is sufficient for the purpose.
Clearly, since the necessity of the resident share by itself fulfils
the no-cloning theorem, authorized sets are {\em not} required to be
mutually overlapping. Thus monotonicity is the only constraint. \hfill \qed
\bigskip

Some corrolories of the theorem are worth noting. First is that the number ($=\lambda-1$) 
of resident shares is strictly less than the number ($\ge N \ge \lambda$) of player shares.
A share $q$ is `important' if there is an unauthorized set
$T$ such that $T \cup \{q\}$ is authorized. From the fact the Theorem uses a
threshold scheme (the $((\lambda,2\lambda-1))$ scheme) in the first layer, it follows that all
the resident shares are important.

As an illustration of the Theorem, we consider the access structure $\Gamma = \{ABC, BD, EFG\}$, 
for which $\lambda=2$. In the first layer, a $((2,3))$ scheme is employed to split $|S\rangle$ into
three shares, with one share designated to the class $C_1 \equiv \{ABC, BD\}$ and the 
other to $C_2 \equiv \{EFG\}$. The last remains with the dealer.
In the second layer, the first share is split-shared among members of $C_1$ according to
a conventional QSS scheme. The second share is split-shared among players of $C_2$
according to a $((3,3))$ scheme. Diagrammatically, this can be depicted as follows.
\begin{equation}
\label{eq:qz0}
((2,3)) \left\{ \begin{array}{ll}
((2,3)) : & \left\{ \begin{array}{ll} 
				((3,3)) : & A, B, C \\
				((2,2)) : & B, D    \\
				|S^{\prime}\rangle
		    \end{array} \right. \\
 ~~ & ~~ \\
((3,3)) : & E, F, G \\
 ~~ & ~~ \\
((1,1)) : &  {\rm dealer}
	   \end{array} \right. 
\end{equation}

Note that given any $\Gamma$, even with non-zero disconnected pieces (i.e.,
the AS graph is not connected), there is a trivial
AQSS by simply adding a common player to all authorized sets, and designating him
to be the dealer: eg., 
$\Gamma = \{ABC, DE, FGH\}$ giving $\Gamma^{\prime} = \{ABCX, DEX, FGHX\}$,
where shares to $X$ would be designated as resident shares. 
Thereby, the structure $\Gamma =
\Gamma^{\prime}|_{\overline{X}}$, which denotes a restriction of 
$\Gamma^{\prime}$ to members other than $X$,  is effectively
realized among the players (not including the dealer). However, this case is excluded as a
valid AQSS because the number of resultant resident shares are non-minimal,
at least according to the recursive scheme outlined above. In all it would
require $3 + 2x$ ($> 3$) shares, where $x$ is the number of instances in which
$X$ appears in a maximal structure $\Gamma^{\prime}_{\max}$
that includes $\Gamma^{\prime}$. More generally, the
requirement is a minimum of $r + (r-1)x \ge r$ resident shares, where $r$ is the number
of authorized sets in $\Gamma$. A better method
is for the dealer to employ a pure state scheme that implements $\Gamma^{\prime}_{\max}$,
retain all shares corresponding to $X$ while discarding all those
corresponding to sets in $\Gamma^{\prime}_{\max}-\Gamma^{\prime}$.
In all this would require $3$ shares, or, in general, $r$ shares.
In contrast, according to the Theorem above, no more than $\lambda-1=2$ resident shares are 
needed. Clearly, in general, $\lambda - 1 < r$.
These considerations suggest that $\lambda(\Gamma)-1$ is the minimal number of resident
shares required to implement a QSS for $\Gamma$. We conjecture that this is indeed the case.

\acknowledgments
SKS thanks Optics Group,
RRI and SIF, IISc for supporting his visit, during which this work was done.

\end{document}